\documentclass[twocolumn,letterpaper,superscriptaddress,showpacs,prl]{revtex4}
\usepackage{graphicx}
\usepackage[usenames]{color}
\usepackage{amsmath}

\begin{document}
\title{Localization of Dirac electrons by Moir\'e patterns in 
graphene bilayers}

\author{Guy \surname{Trambly de Laissardi\`ere}}
\email[corresponding author: ]{guy.trambly@u-cergy.fr}
\affiliation{Laboratoire de Physique Th\'eorique et Mod\'elisation, Universit\'e de Cergy-Pontoise -- CNRS,
F-95302 Cergy-Pontoise Cedex, France.}
\author{Didier \surname{Mayou}}
\affiliation{Institut N\'eel, CNRS -- Universit\'e Joseph Fourier, F-38042 Grenoble, France}

\author{Laurence \surname{Magaud}}
\affiliation{Institut N\'eel, CNRS -- Universit\'e Joseph Fourier, F-38042 Grenoble, France}

\date{\today}

\begin{abstract}
We study the electronic structure of two Dirac electron gazes coupled by a periodic
Hamiltonian such as it appears in rotated graphene bilayers.
{\it Ab initio} and tight-binding approaches are combined and show  that the spatially periodic
coupling between the two Dirac electron gazes  can renormalize strongly their velocity. We investigate in particular  small  
angles of rotation 
and show that the velocity tends to zero in this limit.  The localization is confirmed by an analysis of the eigenstates which
are  localized essentially in the AA zones of the Moir\'e patterns. 
\end{abstract}

\maketitle

\textit{Introduction.}-- 
Graphene is a two dimensional carbon material which takes the form of
a planar  lattice of $sp^{2}$ bonded atoms. Its honeycomb lattice consists in two
 sublattices  which gives a specific  property to the 
wave function, the so-called chirality. The linear dispersion relation close to the charge neutrality point implies
that, in this energy range, the electrons obey an
effective massless Dirac equation \cite{Novoselov04,Zhou06,Wallace47}. 
The properties of  electrons in
graphene, deriving from the Dirac equation, are fundamentally
different from those deriving from  the  Schr\"odinger equation.  In particular the
quantum Hall effect  is quantized with integer plus half values \!\cite{Novoselov,Kim} 
and can even be observed  at room
temperature \!\cite{NovoselovQHERT}. Another example of the unique behavior of
Dirac electrons is the so-called Klein paradox which is  intimately related to the
chirality of their wavefunction. The Klein paradox is the fact that, in a
one-dimensional configuration, a potential barrier is perfectly transparent for
electrons. As a consequence  it is difficult to localize Dirac electrons with an electrostatic potential,  although it can be of great interest to realize this confinement, in particular for the production of elementary devices \!\cite{Geim,Egger}. 

Therefore it is necessary to improve our knowledge of the behavior of Dirac electrons in various types of potentials. For disordered potentials specific behavior related to the chirality of Dirac electrons, such as weak anti-localization effects, have been predicted
theoretically \!\cite{Ando98} and observed in epitaxial graphene \!\cite{Wu,Berger,Heer07,Berger2}. 
Recent STM measurements, performed on
epitaxial graphene, have also confirmed the absence of backscattering related to the Klein paradox  
\!\cite{Brihuega}. 
For periodic
systems several studies have been published concerning  either a simple graphene sheet in
an applied periodic potential or periodic bilayers  \!\cite{Louie,Barbier09}. The bilayer system  
\!\cite{Latil07,Hass_prl08,CastroNeto,Pankratov} has shown
an unexpected and rich behaviour. The well-known Bernal AB stacking leads to massive quasiparticules with
quadratic dispersion close to the Dirac point. Yet it has been recognized recently in the scientific community that another important case is that of rotated graphene bilayers, which are observed for example in epitaxial graphene on the C-terminated face of SiC \cite{Hass_prl08}.  In a rotated bilayer the superposition of the  two rotated honeycomb lattices generates a Moir\'e pattern  with a longer period. 
The two Dirac  electron gases are then coupled by a periodic interaction, with a large supercell, which can restore  a Dirac-like linear dispersion. 
However  
fundamental issues are controversial. In particular  Lopez dos Santos {\it et al.} \!\cite{CastroNeto} have shown,
with a perturbative treatment of the interplane coupling, that the velocity can be renormalized and tend
to decrease in a rotated bilayer with respect to its value in pure graphene. 
In contrary  Shallcross {\it et al.}  \!\cite{Pankratov}  
conclude from {\it ab initio} calculations and
from some general arguments that the velocity is unchanged in rotated graphene bilayers.

In this letter we combine {\it ab initio} and tight-binding (TB)
approaches
and conclude indeed that  there is a
renormalization of the velocity for bilayers as compared to pure graphene.
Most importantly  the velocity tends to zero in the
limit of small twist angles which means that this type of coupling is able to confine the
two Dirac electron gazes. We also show that the electronic wavefunction tends to
localize in regions of the bilayer which are locally similar to the AA arrangement. 
This new localization regime should be observable since angles as small as a fraction of a degree, for which we predict strong localization effects, are found in some rotated multilayers. In particular large Moir\'e patterns have been observed on STM image of graphene multilayers 
on top of SiC C-face \cite{Varchon_prb08} and also on 
graphite \cite{Rong93}. Our work demonstrates also that the perturbative theory of Lopez dos Santos {\it et al.} \!\cite{CastroNeto} is correct for  angles greater than  typically 3 degrees but cannot deal with the remarkable localization phenomenon  that occurs in the limit of small twist angles.

\textit{Moir\'e patterns and geometry of rotated bilayer.}--
Moir\'e Patterns can be obtained in two cases: when two lattices with slightly different 
parameters are superimposed 
or when two identical lattices are rotated by an angle $\theta$ \!\cite{Campanera}. 
The present situation corresponds to the later case.
A commensurate structure can be defined if the rotation changes 
a lattice vector ${\bf V} ~(m,n)$ to
${\bf V}'~(n,m)$ with $n$, $m$ the coordinates with respect to the basis vectors 
$\vec a_1$ $(\sqrt{3}/2,-1/2)$
and $\vec a_2$ $(\sqrt{3}/2,1/2)$. 
The rotation angle is then defined as follows:
\begin{equation}
\cos\theta = \frac{n^2+4nm+m^2}{2(n^2+nm+m^2)}
\label{eq:Comm3}
\end{equation}
and the commensurate cell vectors correspond to :
\begin{equation}
\vec t = {\bf V}' = n \vec a_1 + m \vec a_2
~~{\rm ;}~~
\vec{t'} = -m \vec a_1 + (n+m) \vec a_2.
\label{eq:Comm4}
\end{equation}
The commensurate unit cell contains $N = 4(n^2+nm+m^2)$ atoms.
It is important to notice that $\theta \sim 0$ results in a large cell --large $n$ and $m$-- 
and small $|m-n|$.
Large commensurate cells can be also obtained for large angles 
$\theta \sim 30^{\rm o}$ then $|m-n|$ is large \!\cite{Campanera}.
$\theta = 60^{\rm o}$ is the perfect AB stacking and $\theta$ close to $60^{\rm o}$ is 
obtained for $n=1$ ($m=1$) and large $m$ ($n$).

\begin{figure}
\includegraphics[angle=0, width=7.5cm,clip]{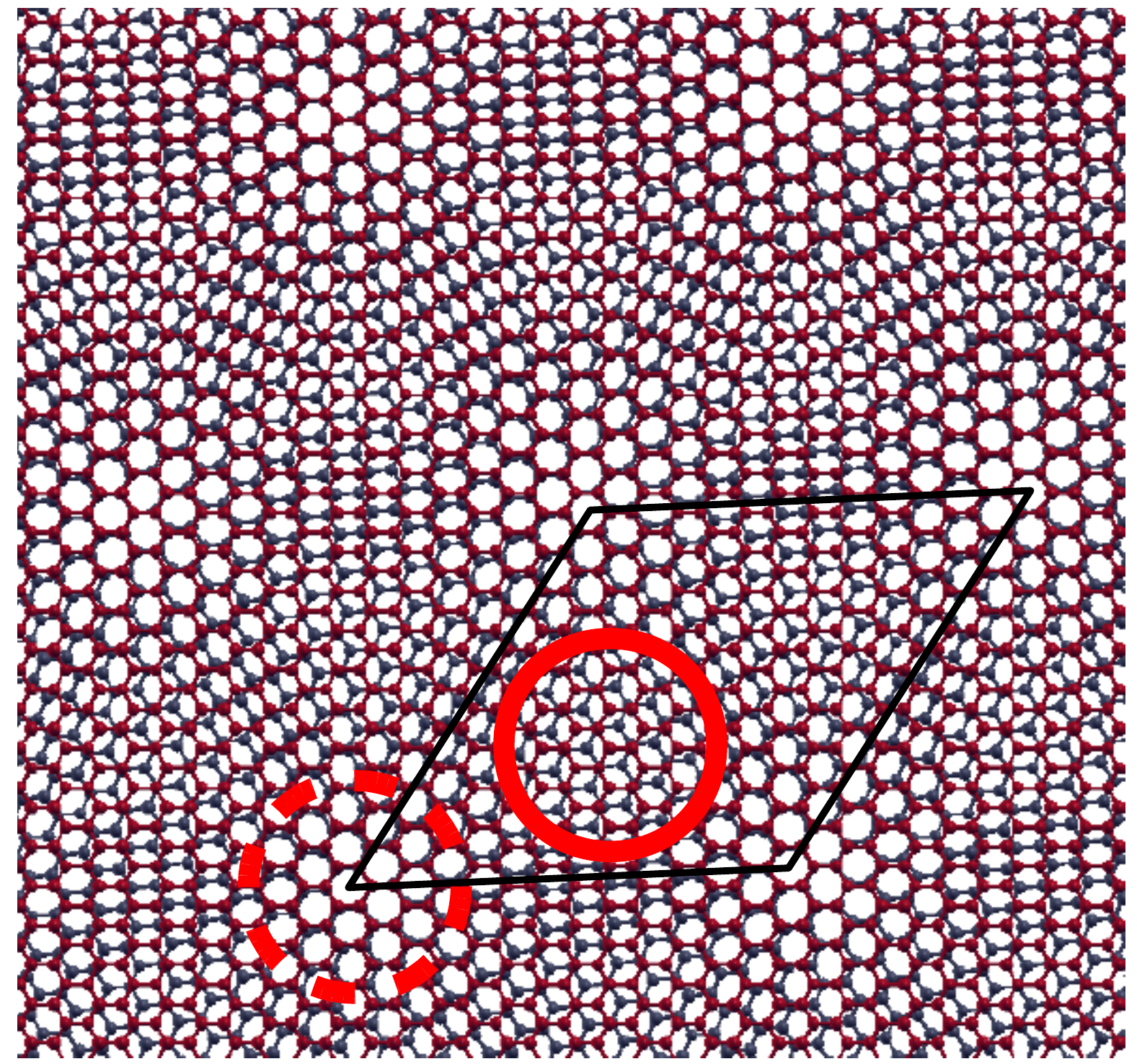}
\caption{(color on line) 
Commensurate bilayer cell $(n,m)=(6,7)$ for a rotation of $\theta = 5.08^{\rm o}$.
Full (dashed) line circle AB (AA) region. 
}
\label{f.1}
\end{figure}

For small angle, 
if the rotation axis, perpendicular to the planes,  passes trough atomic positions  in both layers (figure \ref{f.1}), 
atoms in the 4 corners of the supercell 
are directly superimposed (figure \ref{f.1}). 
This corresponds to the so-called AA stacking.
For $(n,m=n+1)$ cells, zone with AB stacking --graphite like stacking-- are 
located at $1/3$ and $2/3$ of the long diagonal 
(figure \ref{f.1}). 

Quite generally the structures that we study  are fully characterized by the two index $(n,m)$ of the rotation and also by the value of the translation vector between the two layers. However for sufficiently small values of the rotation angle $\theta$ the Moir\'e pattern depends essentially on the rotation angle and the  translation of one layer   only results in a rigid shift of the overall Moir\'e structure. Also in the small angle limit,  all dimensions of the Moir\'e pattern scale like $1/\theta$  and in particular  
the size of locally AA or AB stacked regions 
and the distance between theses zones are proportional 
to $1/\theta$.

Let us emphasize that the limit of small angle is discontinuous: 
when the rotation angle decreases  the dimensions of the Moir\'e structure scale like $1/\theta$ and therefore the structure does not tend to that for  exactly zero angle. This singular, discontinuous, geometric evolution is associated to singular electronic properties as we show in this work.

\textit{Ab-initio and tight-binding methods for electronic structure.}--
Our approach combines {\it ab initio} calculations that can be used for unit cells  containing up to 500-600 atoms and tight-binding (TB) calculations that can be used for unit cells containing up to 15000 atoms or even more.
{\it Ab initio} calculations are performed with the code VASP \!\cite{vasp} and the generalized
gradient approximation \!\cite{pw}.  The C ultra soft pseudopotential \!\cite{uspp} has been extensively tested previously \!\cite{Hass_prl08,Varchon_prb08}. The plane wave basis cutoff is equal to 211 \!eV. 
The empty space width is equal to 21 \!\AA, the interlayer distance is fixed to its experimental 
value ($\simeq 3.35$\,{\rm \AA}) and no atom are allowed to relax for direct comparison with TB calculations.

In the tight-binding scheme only $p_z$ orbitals are taken into account since we are interested in electronic states close to the
Fermi level. Since the planes are rotated, neighbors are not on top of each other (as it
is the case in the Bernal AB stacking). Interlayer interactions are then not restricted
to $pp\pi$ terms but some $pp\sigma$ terms have also to be introduced. 
For both terms, a parameter and a characteristic length have been fitted on 
AA, AB, (1,3) and (1,4) cells \cite{a_paraitre} to reproduce the {\it ab initio} dispersion curves 
according to:
\begin{eqnarray}
V_{pp\pi} &=&-\gamma_0 \exp \left(q_{\pi} \left(1-\frac{d}{a}\right) \right) \\
V_{pp\sigma} &=& \gamma_1 \exp \left(q_{\sigma} \left(1-\frac{d}{a_1} \right) \right)
~~{\rm with}~~\frac{q_{\sigma}}{a_1}=\frac{q_\pi}{a}
\label{eq:tb1}
\end{eqnarray}
$a$ is the nearest neighbor distance within a layer $a=1.42$\,{\rm \AA} and $a_1$ is the
interlayer distance $a_1=3.35$\,{\rm \AA}. 
With this dependence, we take first neighbors interaction in
a plane equal to $\gamma_0 = 2.7$\,eV 
and second neighbors interaction in a plane equal to
$0.1\times\gamma_0$ \cite{Castro09_RevModPhys}, 
which fixes value of the ratio ${q_{\pi}}/{a}$.
For undoped bilayers, all $p_z$ orbitals have the same on-site energy in both planes. We have chosen this on-site energy such that  the energy of the Dirac point equals to $0$ for the results detailed here.

\begin{figure}
\includegraphics[angle=0, width=8.0cm,clip]{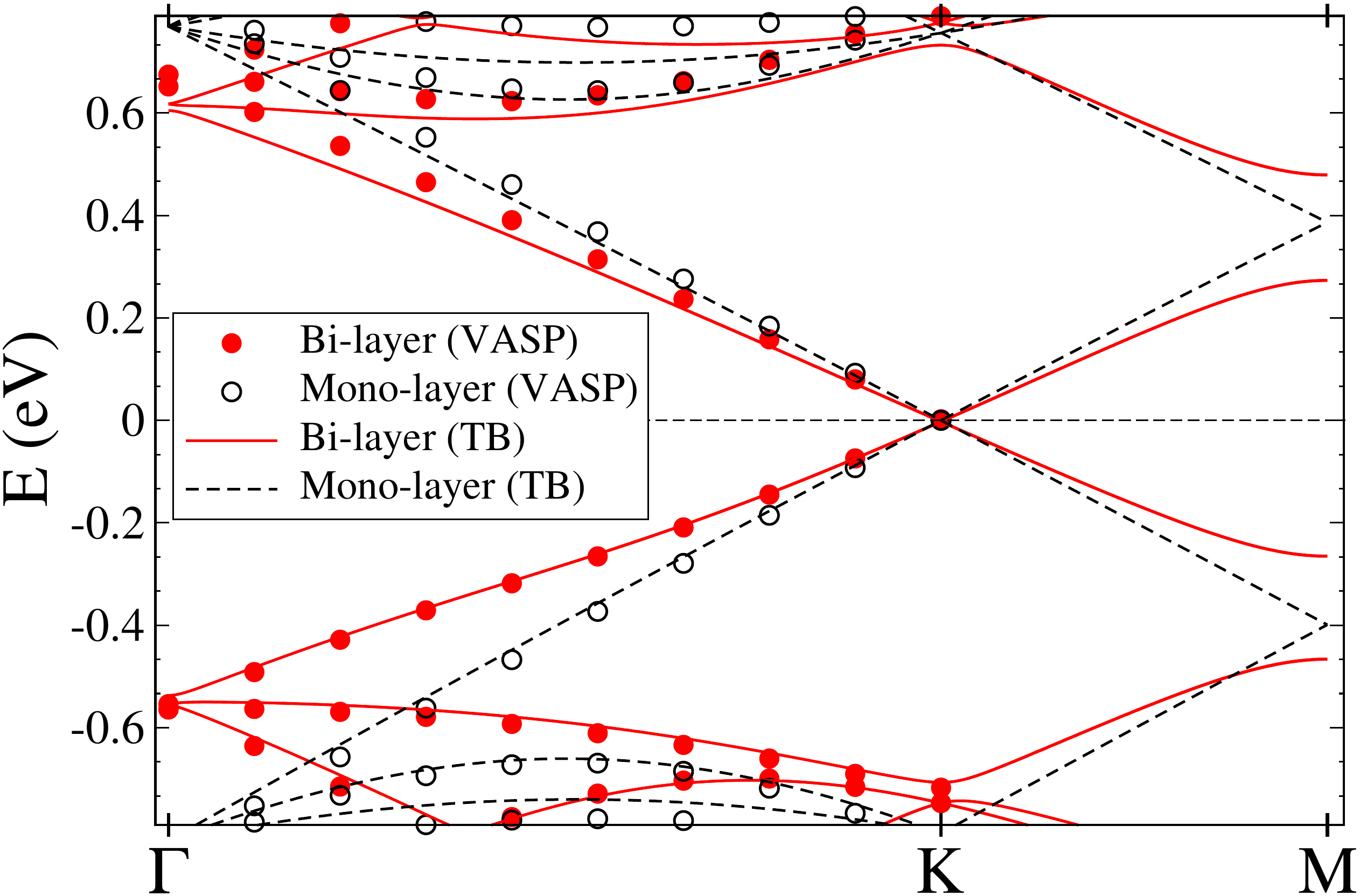}
\caption{(color on line) Band energy dispersion $E(\vec k)$ 
in commensurate bilayer cell $(n,m)=(6,7)$ for a rotation of $\theta = 5.08^{\rm o}$.}
\label{f.2}
\end{figure}

\begin{figure}
  \includegraphics[clip,width=0.45\textwidth]{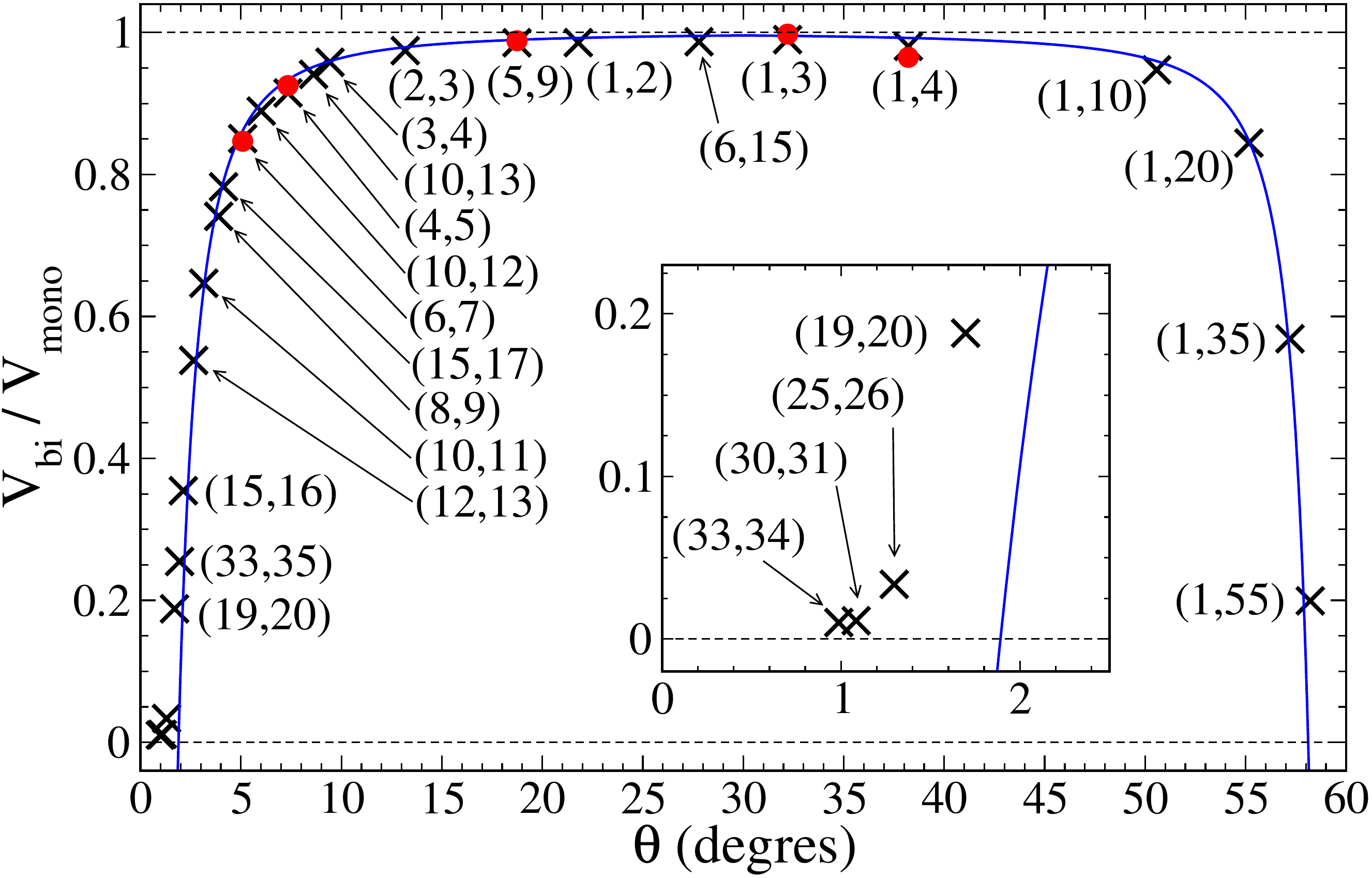}
  \caption{(color on line) Velocity ratio $V_{\rm bi}~/V_{\rm mono}$ 
for a commensurate $(n,m)$ bilayer cell versus rotation angle $\theta$:
Circle VASP, cross TB calculations. 
Line is the model of Lopez dos Santos {\it et al.} [18]: 
$V_{\rm bi}~/V_{\rm mono} = 1-9[\tilde{t}/(V_{\rm mono} K \time 2 \sin(\theta /2))]^2$, with
$\tilde{t} = 0.11$\,eV and $V_{\rm mono} K = 2\gamma_0 \pi \sqrt{3} = 9.8$\,eV.
}
  \label{f.3}
\end{figure}

\textit{Localization by Moir\'e patterns.}--
We have performed band structure calculations on a large number of structures.  For structures that could be studied by both {\it ab initio} and TB methods the agreement is always excellent. For example figure \ref{f.2} presents the band structure calculated with VASP and with the TB method for a small angle rotation:  $(n,m)=(6,7)$, $\theta = 5.08^{\rm o}$ and  $N = 508$. The two methods agree very well and  show a decrease of the velocity by about $15$ percent in that case. We have checked that the velocity remains isotropic in the linear region up to a few percents. For angles in the range $15^{\rm o}-45^{\rm o}$ and for energies, of the order of a few meV (not shown here), the dispersion is not always linear and a very small gap can even exist. Yet this small energy range is not experimentally relevant in most circumstances.  We note also that for angles within a few degrees of $0^{\rm o}$ or  $60^{\rm o}$ the bands become flat and the energy range in which the bands are linear decreases. 

As stated above the geometry of the Moir\'e depends essentially on the angle $\theta$, in the limit of small $\theta$. This reflects in  the band dispersion which depends essentially on the rotation angle $\theta$ between the two layers  for all cases studied here and in particular in the small angle limit. We performed a systematic study of the renormalization of the velocity close to the Dirac point, compared to its value in a monolayer graphene, for rotation angles $\theta$ varying  between $0^{\rm o}$ and $60^{\rm o}$ (figure \ref{f.3}). 

The renormalization of the velocity varies  symmetrically around $\theta= 30^{\rm o}$. Indeed,
the two limit cases $\theta = 0^{\rm o}$ --AA stacking-- 
and $\theta = 60^{\rm o}$ --AB stacking-- 
are different,  but Moir\'e patterns when $\theta \rightarrow 0^{\rm o}$ and when $\theta \rightarrow 60^{\rm o}$
are similar because a simple translation by a vector transforms an AA zone to an AB zone.

Focusing on  angles smaller than $30^{\rm o}$ we define three regimes as a function of the rotation angle $\theta$.
For large $\theta$ ($15 ^{\rm o}\leq \theta \leq 30^{\rm o}$) the Fermi velocity is very close to that of graphene (see figure \ref{f.3}). 
For intermediate  values of $\theta$ ($3 ^{\rm o}\leq \theta \leq 15^{\rm o}$) the perturbative theory of 
Lopez dos Santos {\it et al.} \!\cite{CastroNeto} predicts correctly the velocity renormalization.
But for the small rotation angles ($\theta \leq 3^{\rm o}$) 
a new regime occurs where the velocity tends to zero. This remarkable localization regime cannot be described by the perturbative theory of Lopez dos Santos {\it et al.} ~\cite{CastroNeto}.

\begin{figure}
  \includegraphics[clip,width=0.43\textwidth]{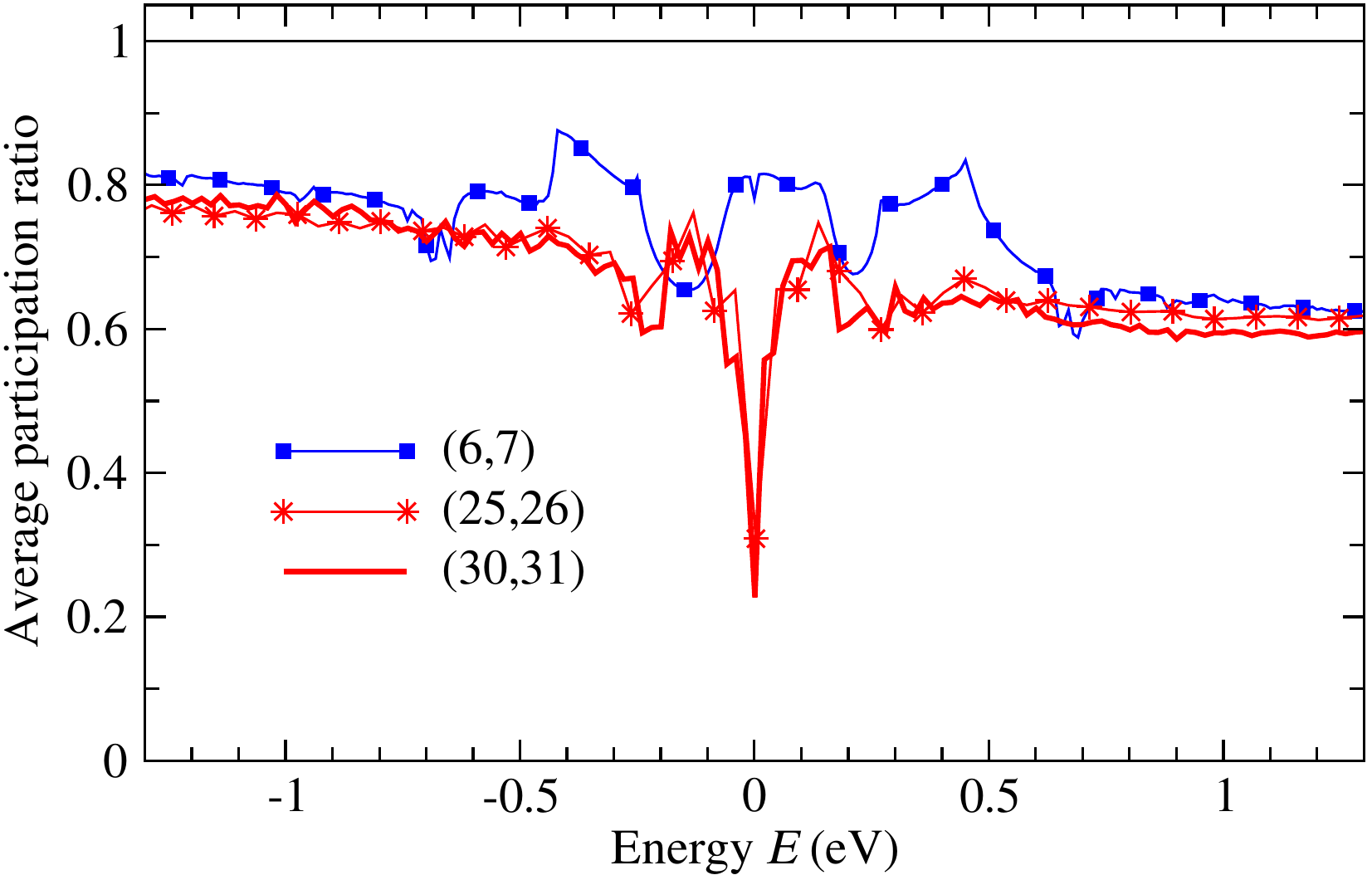}
  \caption{(color on line) Average participation ratio $\tilde{p}(E)$ at energy $E$ for several $(n,m)$ bilayers. 
}
  \label{f.4}
\end{figure}

\begin{figure}
\includegraphics[clip,width=0.45\textwidth]{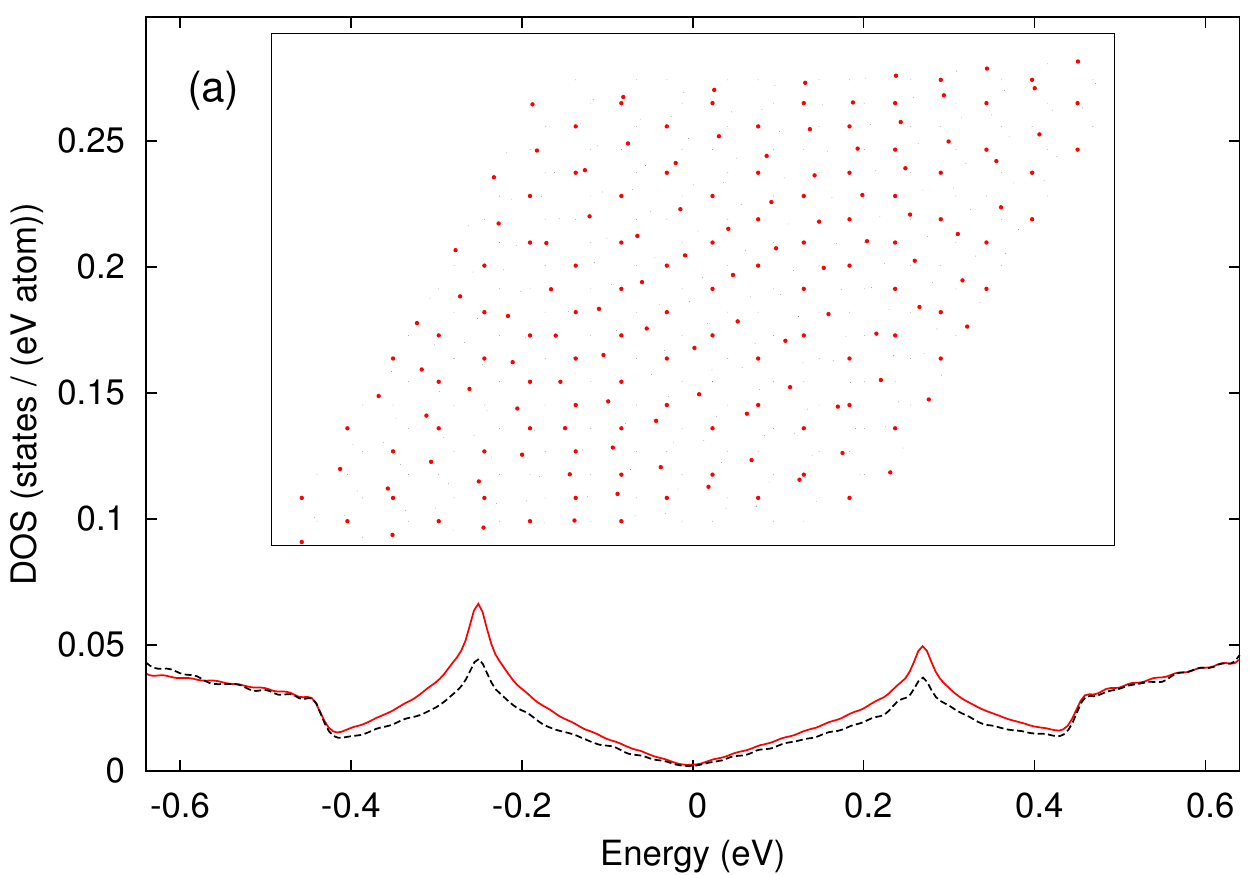}

\includegraphics[clip,width=0.45\textwidth]{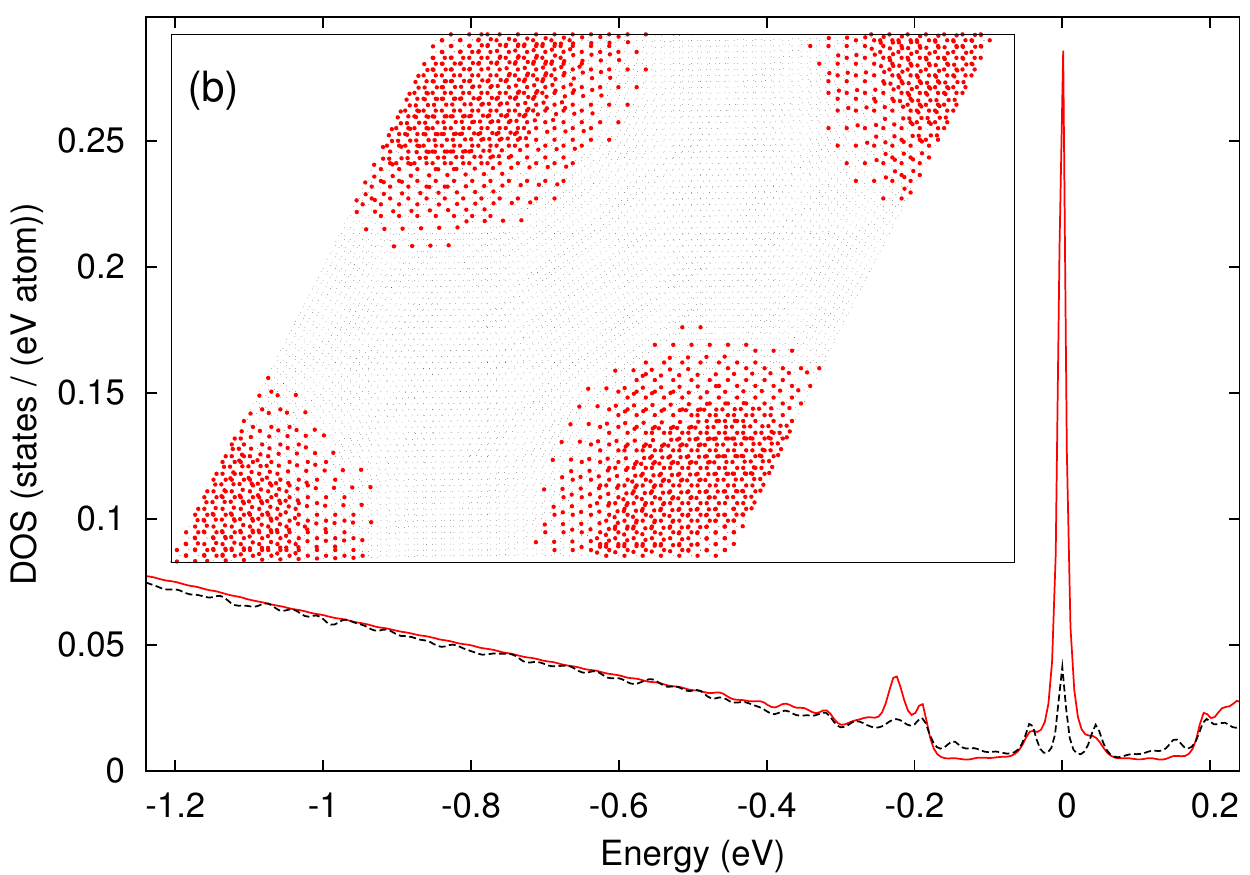}
\caption{(color on line)  TB Local density of states on atom located at the center of a AA type zone 
in (a) $(6,7)$ and (b) $(30,31)$  bilayers. 
Insert: Repartition of a electronic state at {\bf K} with energy $0$: Black small dots are the position of all atoms, red dots are atoms on which 80\% of the electronic state is localized.}
\label{f.5}
\end{figure}

In order to analyze this localization phenomenon we
computed  the participation ratio of each  eigenstate $| \psi \rangle$
defined by $p(\psi) = (N \sum_i |\langle i | \psi \rangle|^{4})^{-1}$.
$| i\rangle$ is the $p_z$ orbital on atom $i$ and $N$ is the number of 
atoms in a unit cell.
For completely delocalized eigenstate $p$ is equal to $1$
like in graphene. On the other hand, state localized on 1 atom have a small 
$p$ value: $p=1/N$.  
The average participation ratio $\tilde{p}$ at energy $E$ is presented on 
figure \ref{f.4}. 
For intermediate  values of $\theta$
--bilayers $(6,7)$  in figure \ref{f.4}--  
the participation ratio of states with energy close to $0$ is similar to that of 
other states of the bilayer. 
For very small $\theta$ 
--bilayer $(25,26)$ and $(30,31)$ in figure \ref{f.4}--
states with energy around $0$ are strongly localized 
(the smallest $p$ values down to $0.14$).
An analysis of spatial repartition of eigenstates, shows that theses states are localized on the AA zones
of the Moir\'e.  This is illustrated on figure \ref{f.5}b showing $80\%$ 
of the weight of an eigenstate  with energy very close to $0$ in the $(30,31)$ bilayer cell.
This new localization mechanism by the potential 
of the Moir\'e leads to a peak in the local 
density of states of atoms in the AA zone (figure \ref{f.5}b) whereas for Moir\'e due to larger angle
no strong localization occurs (figure \ref{f.5}a).
It would be interesting to check whether this localized peak can be observed  in STS experiments.

Finally we have also studied the case of asymmetric bilayers with on-site energies that are different  for atoms of the two planes. Such an asymmetry can occur  in graphene multilayers due to the effect of doping. In the case of Bernal AB stacking  an asymmetric bilayer  presents a gap but our TB calculations show that  it is not the case for asymmetric rotated layers. In agreement with Lopez dos Santos {\it et al.} \!\cite{CastroNeto} we find that a difference between  the on-site energies of the two layers  results in a relative shift  of the Dirac cones. 
The linear dispersion is conserved close to each Dirac point and no gap appears.

\textit{Conclusions.}--
In summary, by combining {\it ab initio} and tight-binding calculations, we
have demonstrated that the dispersion relation stays linear in a rotated bilayer close to the Dirac point. Yet  the velocity of 
electrons is renormalized. This renormalization depends essentially on a single parameter which is the rotation angle $\theta$.
For large $\theta$, ($15 ^{\rm o}\leq \theta \leq 30^{\rm o}$), the Fermi velocity is very close to that of graphene. 
For intermediate  values of $\theta$ ($3 ^{\rm o}\leq \theta \leq 15^{\rm o}$) the perturbative theory of 
Lopez dos Santos {\it et al.} \!\cite{CastroNeto} predicts correctly the velocity renormalization.
But for the small rotation angles ($\theta \leq 3^{\rm o}$), a new regime occurs where the velocity tends to zero. This remarkable localization regime cannot be described by the perturbative theory 
of~\cite{CastroNeto}. 
In this regime electrons with energy close to Dirac point  are localized in AA stacked regions. We believe that this localization regime should be observable since angles as small as a fraction of a degree occur  in some rotated multilayers.

\textit{Acknowledgements.}--
 We thank
C. Berger, 
E. Conrad,
W. de Heer,
P. Mallet,  
F. Varchon, J.~Y.~Veuillen, 
P. First
for fruitful discussions.
Computer time has been granted by
Ciment/Phynum 
and S.I.R (Universit\'e de Cergy-Pontoise).

\end{document}